\documentclass[11pt, a4paper]{article}
\usepackage[english]{babel}
\usepackage[utf8]{inputenc}
\usepackage[T1]{fontenc}
\usepackage{booktabs}
\usepackage{a4wide}
\usepackage{amsmath}
\usepackage{amsfonts}
\usepackage{url}
\usepackage[dvips]{graphicx}
\usepackage{calc}
\usepackage{subfigure}
\usepackage{multirow}
\usepackage{lettrine}
\usepackage{titlesec}
\usepackage{etoolbox}

\apptocmd{\sloppy}{\hbadness 10000\relax}{}{}
\newcommand{\half}[1]{\frac{#1}{2}}
\newcommand{\halb}{\half{1}}

\newcommand{\inv}[1]{\frac{1}{#1}}
\newcommand{\dd}{\:\text{d}}

\newcommand{\kzav}[1]{\left(#1\right)}
\newcommand{\hzav}[1]{\left[#1\right]}

\newcommand{\pd}[2]{\frac{\partial #1}{\partial #2}}

\newcommand{\rf}[1]{(\ref{#1})}

\title{Clausius Equation for Horizons in $F(R)$ Gravity}
\author{
	Bohuslav Matou\v{s} \\
	Department of
	Theoretical Physics and Astrophysics\\
	Faculty of Science, Masaryk University\\
	Kotl\'{a}\v{r}sk\'{a} 2, 611 37, Brno\\
	Czech Republic\\
	E-mail: \texttt{bmatous@mail.muni.cz}\\
	\\
	North-Bohemian Observatory and Planetarium in Teplice\\
	Kopern\'{i}kova 3062, 415 01, Teplice\\
	Czech Republic\\
	}

\begin{document}
\maketitle
\abstract{Covariant Hamiltonian formulation of $F(R)$ gravity is used to find 
variation of surface Hamiltonian which leads to formulation of
temperature and entropy of a horizon. These thermodynamic quantities are then 
used to establish an analogy between Einstein equation and Clausius equation.
The obtained results are compared to those from classical General Relativity.} 

\section{Introduction and Summary}
 The idea of interpreting Einstein equations as a thermodynamic equation originated in work of Jacobson \cite{Jacobson} where he found analogy between Einstein equation and Clausius equation using Rindler horizons and Raychadhauri equation. His method was generalised for $F(R)$ gravitational theory in later works \cite{Chirco}\cite{Eling:2006aw}\cite{Bamba:2009gq} where this gravitational theory was~seen as non-equilibrium thermodynamics and the contribution from new gravity term was interpreted as an addition to entropy which caused a dissipative nature of gravity.

Other approach to study thermodynamic aspects of gravity was introduced in works of Padmanabhan and his colleagues \cite{Padmanabhan:2002sha}\cite{Padmanabhan:2002ma}\cite{Padmanabhan:2009jb}. Their approach focuses strongly on Lagrangian and variation principle. It is known that Lagrangian for general relativity can be separated into two parts, the quadratic bulk term and a surface term. When deriving the equations of motion with fixed variation of metric tensor and its conjugated momentum, the surface term vanishes. But in general it can be shown that there is a relationship between the two parts of Lagrangian \cite{Mukhopadhyay}\cite{Kolekar:2010dm} and that both of them contain information about space-time configuration. This sparked interest to the surface Lagrangian which was until then eclipsed by the bulk Lagrangian. It was found in \cite{Padmanabhan:2004fq} that action of the surface Lagrangian on a horizon leads directly to the temperature and the entropy of such horizon. This has put the surface Lagrangian into the spotlight of thermodynamic research of gravity and this surface Lagrangian has been since then thoroughly examined. 

The description of gravity is mostly done in Lagrangian formalism that is caused by the principle of covariance. If one wants to switch from the Lagrangian formalism to Hamiltonian formalism using classical Hamiltonian theory the covariance gets broken because the momentum is defined by partial derivative of a Lagrangian with respect to time derivative of metric. To preserve the covariance, a covariant Hamiltonian formulation of field theory has to be used. Weyl-De~Donder theory \cite{DeDonder}\cite{Weyl}\cite{Struckmeier:2008zz} is such theory. The Weyl-De Donder theory defines the momentum as a partial derivative of a Lagrangian with respect to each partial derivative of each coordinate. That basically means that the momenta have dimensions one higher than the coordinates.  

Covariant Hamiltonian formulation of General Relativity was presented by Hořava \cite{Horava:1990ba} and Parattu \cite{Parattu:2013gwa}. In the latter, it was shown that the situation simplifies significantly when the role of coordinate is not acted by simple metric but a slightly different coordinate $f^{ik} = \sqrt{-g} g^{ik}$. This coordinate is not a new one but it has historical significance \cite{Eddington}.

This article works  with the $F(R)$ theory  and aims to find Clausius equations on horizons. It is following the Padmanabhan's approach and examines the surface Hamiltonians of horizons using method presented in \cite{Parattu:2013gwa}. To do so, covariant formulation of $F(R)$ gravity presented in \cite{Kluson:2020tzn} is used. Albeit the surface Lagrangians in General Relativity and $F(R)$-Gravity are similar when written with connection, they are different when using canonical coordinates. The surface Lagrangian is used to formulate surface Hamiltonian. From the surface Hamiltonian it is visible, that the only thermodynamic difference to General Relativity is the extra factor $F'$ in entropy. This result is expected, but it validates the used approach. 

In the second part, two metrics --- spherically symmetric and general static --- are studied to calculate $T\delta S$ from the surface Hamiltonian. Equations of motion allow to express this relation in a form, which can be then interpreted in a form of Clausius equation. Obtained Clausius equations contain a new member. This member is traditionally interpreted as internal entropy of gravity. However in the used approach, it seems more natural to explain it as a new pressure induced by higher curvature terms of $F(R)$. Interesting is also the fact that the terms containing second derivatives of $F'$ vanish on horizon for both metrics. It remains an open question, whether such behaviour is general or they can contribute to a new member of Clausius equation for some metrics.

This paper uses the East Coast convention with metric signature ($-,+,+,+$) and Latin indices running over 0..3 interval while the Greek ones over 1..3. The fundamental constants $c, G, \hbar, k_B$ are treated as equal to one. The paper is organized in the following manner: second section of this article reviews the covariant formulation of $F(R)$ gravity, third section introduces the method used to examine the horizons. Fourth and fifth sections deals with the example metrics and provides their thermodynamic results.

\section{Review of Covariant Formulation of $F(R)$ Gravity}
This section briefly reviews covariant formulation of $F(R)$-Gravity; the full description can be found in \cite{Kluson:2020tzn}. 
In $F(R)$-Gravity theory, it is an advantage to introduce new two scalar fields $A$ and $B$ and write the Lagrangian density as
\begin{equation}
 \mathcal{L} = \inv{16\pi}\sqrt{-g} \hzav{F(B) + A(R - B)}\:,
\end{equation}
the equation of motion for $B$ leads to relation $A = F'(B)$ where prime denotes the derivative with respect to $B$. This equation of motion can be used to simplify the Lagrangian density by removing the influence of scalar field $A$ completely, leaving us with the Lagrangian density
\begin{equation}
 \mathcal{L} = \inv{16\pi}\sqrt{-g} \hzav{F(B) + F'(R - B)}\:.
\end{equation}
Solving the equations of motion yields
\begin{align}
G_{ik} &= \frac{8\pi}{F'} \kzav{T_{ik} + \mathcal{T}_{ik}} \:, \\
\mathcal{T}_{ik} &= \inv{8\pi} \hzav{
\halb \kzav{F - F' B} g_{ik} + \nabla_k \nabla_i F' - \Box F' g_{ik} } \:, \\
0 &= F'' (R - B) \:,
\end{align}
where $\Box = \nabla^i \nabla_i$. Let us focus on the last equation, it leads to $R = B$ with condition $F''(B) \neq 0$. Breaking on this condition leads to standard General Relativity. The next step is to split the Lagrangian density into two parts: a bulk term and a surface term. The bulk term is
\begin{align}
 &16\pi\mathcal{L}_{bulk} = 
  \sqrt{-g} F'(B) \kzav{g^{ba} \Gamma^c_{ad} \Gamma^d_{bc} - g^{bc} \Gamma^a_{ad} \Gamma^d_{bc}} - \\
 &- \sqrt{-g} F''(B)\kzav{\partial_c B} \kzav{g^{bd}\Gamma^c_{bd} - g^{bc} \Gamma^d_{bd}}  + 
  \sqrt{-g} \hzav{F(B) - F'(B)B} \nonumber\:.
\end{align}
The density of surface Lagrangian, which is the more important one for the analysis of thermodynamics of horizons, has the form of
\begin{equation}
\label{L_sur_raw}
 \mathcal{L}_{sur} = \inv{16\pi}\partial_c \hzav{\sqrt{-g} F'(B) \kzav{g^{bd} \Gamma^c_{bd} - g^{bc} \Gamma^d_{bd}}} \:.
\end{equation}
The bulk term might be then used for formulation of covariant Hamiltonian. To do so, the coordinate is changed and instead of metric tensor, tensor 
$f^{ik}$, where $f^{ik} = \sqrt{-g} g^{ik}$, is used. Momentum conjugate to $f^{ik}$ is
\begin{eqnarray}
 N^c_{ik} = 
 - \frac{F'(B)}{16\pi} \Gamma^c_{ik}
 + \frac{F'(B)}{32\pi} \kzav{\Gamma^d_{id} \delta^c_k + \Gamma^d_{kd} \delta^c_i} +\\
 + \frac{F''(B)}{32\pi} \kzav{\partial_i B \delta^c_k + \partial_k B \delta^c_i} 
 + \frac{F''(B)}{32\pi} \partial_d B f^{dc} f_{ik} \nonumber \:,
\end{eqnarray}
and momentum conjugate to $B$ is 
\begin{equation}
\label{p^n}
 p^n = \frac{F''}{16\pi} \kzav{\Gamma^b_{bd} f^{dn} - \Gamma^n_{bd} f^{bd}} \:,
\end{equation}
which can be used to express surface Lagrangian density \eqref{L_sur_raw} as
\begin{equation}
\label{L_sur}
\mathcal{L}_{sur} = -\partial_c \kzav{\frac{F'}{F''}p^c} \:.
\end{equation}
Which we can compare to the surface Lagrangian of General Relativity
\begin{equation}
\mathcal{L}_{sur} = -\partial_c \kzav{f^{ab} N^c_{ab}} \:,
\end{equation}
It is interesting that the surface Lagrangian of $F(R)$-Gravity uses only new coordinates $B$ and $p^a$ and is independent of $f^{ab}$ and $N^c_{ab}$. Such difference seems to be tremendous but when written with the connection there is only one extra $F'$ in the surface Lagrangian of $F(R)$-Gravity.

\section{Surface Hamiltonian}
Let us now examine the surface Lagrangian and its relations to boundaries. The basic relationship for us is the Hamilton's principle of least action which tell us
\begin{equation}
 \mathcal{A} = \int_\mathcal{M} \mathcal{L} \dd^4 x \:,
\end{equation}
where $\mathcal{M}$ is a manifold and $\mathcal{A}$ is the action. Since the action is functional, which is linear in Lagrangian, it can be separated in the very same manner as the Lagrangian and through it we obtain two actions --- bulk and surface
\begin{equation}
 \mathcal{A} = \mathcal{A}_{bulk} + \mathcal{A}_{sur\!f}  = \int_\mathcal{M} \mathcal{L}_{bulk} \dd^4 x  + \int_\mathcal{M} \mathcal{L}_{sur\!f} \dd^4 x \:.
\end{equation}
The surface Lagrangian, as its name advises, is significant only on boundaries, because it has a form of divergence of some vector potential
\begin{equation}
 \mathcal{L}_{sur\!f} = \partial_i \kzav{\sqrt{-g}V^i} \:.
\end{equation}
Thus, the surface action is
\begin{equation}
 \mathcal{A}_{sur\!f}  = 
 \int_\mathcal{M} \mathcal{L}_{sur\!f} \dd^4 x = 
 \int_\mathcal{M} \partial_i\kzav{\sqrt{-g}V^i} \dd^4 x =
 \oint_{\partial \mathcal{M}} \sqrt{-g}V^i \dd B_i
 \:.
\end{equation}
where $B_i$ is the unit covector normal to the manifold's border $\partial M$.
This is the point, where we put in a bit of thermodynamics and use formula, which defines surface Hamiltonian and relates surface action to entropy \cite{Parattu:2013gwa}\cite{Padmanabhan:2012qz}
\begin{equation}\label{H_surf}
 \mathcal{H}_{sur\!f} = - \pd{\mathcal{A}_{sur\!f}}{\tau} = TS \:,
\end{equation}
where $\tau$ is the proper time, $T$ is temperature of the horizon and $S$ its entropy, the minus sign is a matter of convention, which we will follow. 

Note, that the total Hamiltonian can be also separated same way as Lagrangian and action and thus
\begin{equation}
 \mathcal{H}_{total} = \mathcal{H}_{sur\!f} + \mathcal{H} = TS + \mathcal{H} \:,
\end{equation}
which reminds us the Legendre transformation of thermodynamic potentials. Therefore if say that the total Hamiltonian $\mathcal{H}_{total}$ is an analogy of inner energy then the bulk Hamiltonian can be seen as analogy of free energy \cite{Padmanabhan:2004fq}.

However the proper time in \rf{H_surf} makes it difficult to find general solutions, but if we restrict ourselves to static metrics, we can use the following formula
\begin{equation}\label{H_surf_static}
 \mathcal{H}_{sur\!f}  = 
 -\oint_{H} \sqrt{-g}V^n \dd H_n
 \:,
\end{equation}
where $H$ is the two dimensional surface of a horizon and $H_n$ is direction normal to the surface.

Vector potential $V^n$ for $F(R)$-Gravity is from \rf{L_sur_raw}
\begin{equation} \label{def_V}
V^n = \inv{16\pi} F' \kzav{g^{ab} \Gamma^n_{ab} - g^{na} \Gamma^b_{ab}} \:,
\end{equation}
let us compare it with the vector potential of General Relativity
\begin{equation}
 ^{GR}V^n = \inv{16\pi}  \kzav{g^{ab} \Gamma^n_{ab} - g^{na} \Gamma^b_{ab}} \:,
\end{equation}
the only difference is the factor $F'$. Since the surface Hamiltonian is linear in $V^n$, the only difference between surface Hamiltonians for General Relativity and $F(R)$-Gravity will also be factor of $F'$. From this we can see that the temperature relation will be the same and entropy shall take this extra $F'$.\footnote{This addition to entropy is expected and serves as a validation that the used method yields correct results. When using the entropy tensor $P^{ab}_{cd} = \pd{\mathcal{L}}{R^{cd}_{ab}}$ \cite{Padmanabhan:2012qz} it is easily visible that for $F(R)$-Gravity it takes form of $P^{ab}_{cd} = F' \left. P^{ab}_{cd}\right|_{GR}$ so the entropy tensor has the same factor $F'$ comparing to General Relativity.}

\section{Spherical case}
Having formulated the general surface Hamiltonian, it is time to look for Clausius equation. To do so, a metric needs to be put in. The simplest relevant case is the spherically symmetric metric. We will use it to find relations for temperature and entropy. Then the $T\delta S$ part of variation if the surface Hamiltonian is taken and simplified using equations of motion to find a form, which is then subject of thermodynamic interpretation. 

The spherically symmetric metric is taken in the following form 
\begin{equation}
 \dd s^2 = -a \dd t^2 + \inv{a} \dd r^2 + r^2 \dd \vartheta^2 + r^2 \sin^2 \vartheta \dd \phi^2\:,
\end{equation}
where $a$ is a function of $r$ which satisfies conditions $a(r_h) = 0$ and $a'(r_h) = 2 \kappa \neq 0$ for a parameter $r_h$ and surface gravity $\kappa$. The normal coordinate in this case is radius  $r$. Vector potential $V^i$ can be easily calculated using \eqref{def_V} as
\begin{equation}
 V^1 = \frac{F'}{16\pi} \kzav{-\frac{4 a}{r} - a'} \:.
\end{equation}
Having calculated this potential, we can put it into \eqref{H_surf_static} and get variation of surface Hamiltonian in form 
\begin{equation}
\delta H_{sur} = \left.
\inv{16\pi}\delta\int \dd \vartheta \int \dd \phi 
 \sin \vartheta
 \hzav{
 F' \kzav{4ar + a' r^2}}\right|_{r = r_h}\:.
\end{equation}
After the integration, it can be modified to the simplified form using the formula for surface area $A = 4\pi r_h^2$ 
\begin{equation}
\label{dH_final}
 \delta H_{sur} = \delta \hzav{
\kzav{\frac{\kappa}{2\pi}}
\kzav{\frac{F'_h A}{4}}
     }\:,
\end{equation}
which is equivalent to the relationship $\delta E = T \delta S + S \delta T$ and where we have used $F'_h$ for the value of $F'$ on $r_h$. In the spherical symmetric case, we can use that such value is constant because $F$ is a function of scalar curvature, which is a function of radius only. From \eqref{dH_final}, it is clearly visible that the GR relation for temperature $T = \frac{\kappa}{2\pi}$ holds for the $F(R)$ gravity case. While the definition of entropy $S = F' \frac{A}{4}$ differs only in factor $F'$. This factor goes to one in the GR limit. This result concords with our expectations from previous section as well as with the GR result from \cite{Parattu:2013gwa}.

To move further, Einstein equations on the horizon need to be evaluated. The Einstein tensor $G^i_k$ takes the GR form
\begin{equation}
 G^0_0 = G^1_1 = \frac{a'r + a - 1}{r^2}\:, 
 \qquad 
 G^2_2 = G^3_3 = \frac{2a' + a''r}{2r}\:.
\end{equation}
For simplicity let's assume that tensor $T^i_k$ takes the form of perfect fluid. The extra member of Einstein equations for $F(R)$-Gravity can be divided into two part, the first one, which cannot be further simplified without knowing $F(R)$, and the part depending on the second derivative of $F'$, which needs our attention. In the process of calculating of $\nabla^a \nabla_b F'$, we shall make use of the second equation of motion $B = R$ and take $F$ as a function of scalar curvature $R$ directly. We can also make use of spherical symmetry where the scalar curvature is a function of the radius only. Being interested only in the equations of motion on horizon allows us to neglect terms containing solely $a$ since it vanishes on the surface. Having taken all this into consideration, we get
\begin{equation}
\left. \nabla^a \nabla_b F'\right|_{r=r_h} = 
- \left.\kzav{ g^{ac} \Gamma^1_{cb} F'' \partial_1 R }\right|_{r=r_h}
\:,
\end{equation}
next step is substitution of the Christoffel symbols and evaluation of Einstein equations on the horizon
\begin{align}
 &\frac{2\kappa r_h- 1}{r_h^2} =
 \frac{8\pi }{F'_h} T^1_1 +
 \frac{F_h - F'_h R_h}{2F'_h} -
 \frac{F''_h}{F'_h} \kappa R^{(1)}_h  \:, \\
 &\frac{2\kappa+ \kappa' r_h}{r_h} =
 \frac{8\pi }{F'_h} T^1_1 +
  \frac{F_h - F'_h R_h}{2F'_h} -
 2 \frac{F''_h}{F'_h} \kappa R^{(1)}_h\:,
\end{align}
where $F|_{r=r_h} = F_h$, $F''|_{r=r_h} = F''_h$, $R|_{r=r_h} = R_h$, and $\kzav{\partial_1 R}|_{r=r_h} = R^{(1)}_h$.
The first equation can be used for expressing $\kappa$ as
\begin{equation}\label{kappa_sphere}
 \kappa  = \inv{2r_h} + \frac{4\pi }{F'_h} T^1_1 r_h + \frac{F_h - F'_h R_h}{4 F'_h} r_h 
 - \halb \frac{F''_h}{F'_h} \kappa r_h R^{(1)}_h \:.
\end{equation}
Now we take $T\delta S$ from the equation \eqref{dH_final}
\begin{equation}
T\delta S = \frac{\kappa}{2\pi}\delta\kzav{F'_h \pi r_h^2} \:,
\end{equation}
 and substitute $\kappa$ from \eqref{kappa_sphere} yields
 \begin{equation}
   T\delta S = \kappa \frac{ r_h^2}{2} \delta F'_h +
   \frac{F'_h\delta  r_h}{2}  +
   4\pi T^1_1 r_h^2 \delta  r_h  + 
   \frac{F_h - F'_h R_h}{4} r_h^2 \delta  r_h  -
  \kappa\frac{r_h^2}{2} F''_h   R^{(1)}_h \delta r_h \:.
 \end{equation}
 Using the fact that $\delta F'_h = F''_h R^{(1)}_h \delta r_h$ the first member and the last member cancel each other. Leaving the equation in the form of
 \begin{equation}
   T\delta S = 
   \frac{F'_h\delta  r_h}{2}  +
   4\pi T^1_1 r_h^2 \delta  r_h  + 
   \frac{F_h - F'_h R_h}{4} r_h^2 \delta  r_h  \: .
 \end{equation}
 This is the point where the members on the right hand side shall be interpreted as thermodynamic quantities. The Missner-Sharp energy for the horizon can be found in the first order as $E= \half{r_hF'}$ \cite{Cai:2009qf}. Thus, the first term on the right hand side can be interpreted as variation of energy. Comparing it to the GR result, it is visible that it differs only in the factor $F'$ which goes to one in the GR limit. The second member on the right hand side contains a variation of volume and it can be easily interpreted as $p_M \delta V$ member, where $T^1_1 = p_M$, which is exactly the same as in the GR case. The only additional member to the GR case is the last one on the right hand side. There are two possible ways of its interpretation. The first one is to notice the similarity with a pressure member volume and interpret it as an another pressure using
 \begin{equation}
 \frac{F_h - F'_h R_h}{4} r_h^2 \delta  r_h = 
 \frac{F_h - F'_h R_h}{16\pi} \delta  V =
 p_\Lambda \delta  V \:,
 \end{equation}
where a new pressure was introduced
 \begin{equation}
  p_\Lambda =  \frac{F_h - F'_h R_h}{16\pi}\:,
 \end{equation}
 if the function $F(R)$ is taken as $F(R) = R - 2\Lambda$, then this new pressure would take form of $p_\Lambda =  -\frac{\Lambda}{8\pi}$
\footnote{One could of course argue, that taking $F(R) = R - 2\Lambda$ breaks on condition $F''\neq 0$ we have been using all along. But then we can just assume that $\Lambda$ is a function and will obtain the new pressure as $p_\Lambda = -\frac{\Lambda  - R \Lambda'}{8\pi}$.}. This allows us to see it as a kind of generalization of dark energy pressure. 
 
 The second interpretation was done in \cite{Akbar:2006mq} where the spherical symmetric metric in $F(R)$ gravity was also analysed using a different approach. They interpreted the last member as an additional internal entropy
 \begin{equation}
  T \delta S^i = 
  -\frac{F_h - F'_h R_h}{4} r_h^2 \delta  r_h = 
  T  \frac{A}{4} \frac{ F'_h R_h - F_h}{2\kappa} \delta  r_h\:.
 \end{equation}

 Having analysed spherically symmetric case, we have found horizon temperature and entropy. The temperature was the same as in the GR case and the entropy differs only in factor of $F'$. Then the thermodynamic equation was looked for yielding equation with one new extra member compared to the GR case. Two possible interpretations were formulated for this equation
  \begin{equation}
  T\delta S =   \delta E  +  p_M \delta V + p_\Lambda \delta V
  \:, \quad
  T\delta S =   \delta E  +  p_M \delta V - T\delta S^i \:,
 \end{equation}
where in the first one, the extra member was interpreted as a pressure arising from the function $F$. The second possible interpretation is as a new entropy which depends not only on horizon surface but also on the function $F(R)$ and the surface gravity.

\section{General static spacetime}
In this section same procedure is performed as in the previous. Now, the examined metric is the general static space-time. As in the previous case, the metric is static, however there is general two dimensional metric on horizon. Explicitly, we have
\begin{equation}
\dd s^2 = -N^2 \dd t^2 + \dd n^2 + \sigma_{AB} \dd x^A \dd x^B \:,
\end{equation}
where major Latin indices $A, B, ...$ denote the indices of the horizon and take value 2 or 3. The coordinate $n$ corresponds to the normal direction to the horizon. The horizon is expected at the point $n = 0$. Near the horizon, we expect the following expansions in $n$ \cite{Medved:2004ih}
\begin{equation}
 N = \kappa n + \mathcal{O}(n^3)\:, \qquad
 \sigma_{AB} = \overline{\sigma}_{AB} + \halb \widehat{\sigma}_{AB} n^2  + \mathcal{O}(n^3)\:,
\end{equation}
where $\kappa$ is the surface gravity of the horizon. To formulate the variation of surface Hamiltonian, the first step is to use the metric to find vector potential $V^n$ using \eqref{def_V}
\begin{equation}
 V^1 = - \frac{F'}{16\pi} \kzav{
 \frac{2\dot{N}}{N} + \sigma{AB}\dot{\sigma}_{AB}
} 
\:,
\end{equation}
where dot indicates derivative with respect to the normal coordinate $n$.
Then we can formulate the surface Hamiltonian as\footnote{For our purposes, it is not important but the expression in the bracket can also be written as $\kzav{2N\sqrt{\sigma}}\dot{}$.}
\begin{equation}
\mathcal{H}_{surf} = \inv{16\pi} \int \dd^2 x_\parallel 
\left.\hzav{
F' \kzav{2 \dot{N} \sqrt{\sigma} + N \sqrt{\sigma} \sigma^{AB} \dot{\sigma}_{AB}}
}\right|_{n \rightarrow 0} \:,
\end{equation} 
using the limit on horizon with $\overline{\sigma}$ being the limit of $\sigma$ on horizon leads to
\begin{equation}
\mathcal{H}_{surf} = \inv{16\pi} \int \dd^2 x_\parallel 
2 \kappa F'_h  \sqrt{\overline{\sigma}} 
\:.
\end{equation} 
Now let us perform the integration using formula for surface area $A = \int \sqrt{\sigma} \dd^2 x_\parallel$ and for the sake of simplicity let us assume that $F'_h$ is constant along the horizon. The variation of surface Hamiltonian is then
\begin{equation}
 \delta H_{sur} = \delta\hzav{
 \kzav{\frac{\kappa}{2\pi}}  \kzav{ \frac{F'A}{4}}
}
\:.
\end{equation}
When interpreting the variation as $\delta H = T\delta S + S \delta T$, the result is identical to the one derived in the spherical case. Again as in the spherical symmetric case, the temperature remains same as in GR while the entropy takes a factor of $F'$. For general case, when we do not want to restrict $F$ we shall use formulation before integration
\begin{equation}\label{H_sur_gss_integral}
\delta \mathcal{H}_{surf} = \inv{16\pi} \int \dd^2 x_\parallel \hzav{
2F'_h\sqrt{\overline{\sigma}}\delta\kappa
+ 2\kappa \delta \kzav{F'_h \sqrt{\overline{\sigma}}}
}\:.
\end{equation}

To formulate the thermodynamic equation on the horizon, Einstein equations have to be examined. Focusing on coordinates $n$ and $t$, it is easy to find members of Einstein tensor. Non-zero are only the diagonal terms $G^0_0$ and $G^1_1$. Albeit they differ in general, when doing limit on the horizon both take the same form \cite{Medved:2004ih}
\begin{equation}
G^0_0\Big|_H = \halb \sigma_2 - \halb \overline{R}_\parallel =           G^1_1\Big|_H \:,
\end{equation}
where  $\sigma_2 = \sigma^{AB}\widehat{\sigma}_{AB}$ and $R_\parallel$ is a Ricci scalar calculated using the two-dimensional metric $\sigma$. Since the method of its calculation preserves behaviour with respect to coordinate $n$, it is reasonable to expect that near the horizon $R_\parallel$ behaves as $R_\parallel = \overline{R}_\parallel + \halb \widehat{R}_\parallel n^2$. 

To obtain the right hand side of Einstein equations, second derivative of $F'$ has to be calculated. The calculation is more complicated than in the spherical case since the scalar curvature now depends on normal coordinate as well as both horizontal coordinates. Important is also the form of Ricci scalar curvature 
\begin{equation}
 R=  
 -2\frac{\ddot{N}}{N} 
 -\frac{\dot{N}}{N} \sigma^{AB} \dot{\sigma}_{AB} 
  -\sigma^{AB} \ddot{\sigma}_{AB} 
  + \frac{3}{4} \sigma^{AC} \sigma^{BD} \dot{\sigma}_{AB} \dot{\sigma}_{CD} 
    - \frac{1}{4} \sigma^{AC} \sigma^{BD} \dot{\sigma}_{AC} \dot{\sigma}_{BD} 
    + R_\parallel \:,
\end{equation}
with the following expansion in $n$
\begin{equation}
 R \approx 
 -2\sigma_2 
 + \overline{R}_\parallel 
 - \inv{4} \sigma_2^2 n^2 
 + \frac{3}{4} \sigma^{AB}\sigma^{CD}\widehat{\sigma}_{AD}\widehat{\sigma}_{BC}n^2 
 + \halb \widehat{R}_\parallel n^2 \:.
\end{equation}
This expansion provides us with important result, that $\partial_1 R$ is vanishing on the horizon. Then the general form of both diagonal terms can be calculated. Similarly to the Einstein tensor, the both members are different, however they limit to the same value on the horizon
\begin{align}
 \mathcal{T}^0_0\Big|_H  = \mathcal{T}^1_1\Big|_H &= \inv{8\pi}\left[
 \frac{F_h - F'_h \kzav{\overline{R}_\parallel - 2 \sigma_2}}{2} - \right. \nonumber \\ 
 & -  \left. D^A \hzav{
     F'' D_A \kzav{\overline{R}_\parallel - 2 \sigma_2}
 } 
 + F''\kzav{
     \inv{2} \sigma_2^2 
     - \frac{3}{2} \sigma_3 
     -  \widehat{R}_\parallel
  }\right] \:, \nonumber \\
\end{align}
where $D$ denotes covariant derivative with respect to the two-dimensional metric $\sigma$ and $\sigma_3 = \sigma^{AB}\sigma^{CD}\widehat{\sigma}_{AD}\widehat{\sigma}_{BC}$.
On the horizon, the Einstein equation for $nn$ and $tt$ components are
\begin{align}
\label{Einstein_gss}
 \halb \sigma_2 - \halb \overline{R}_\parallel &= \frac{8\pi}{F'_h} T^1_1 
 + \frac{F_h - F'_h \kzav{\overline{R}_\parallel - 2 \sigma_2}}{2F'_h} - \nonumber \\
 &- \inv{F'_h} D^A \hzav{
     F''_h D_A \kzav{\overline{R}_\parallel - 2 \sigma_2}
 } 
 + \frac{F''_h}{F'_h}\kzav{
     \inv{2} \sigma_2^2 
     - \frac{3}{2} \sigma_3 
     -  \widehat{R}_\parallel } \: . \nonumber \\
\end{align}
To be able to analyse $T\delta S$, variation of $\sqrt{\sigma}$ is needed. To find it, the method from \cite{Parattu:2013gwa} is used. A new parameter $\Lambda$ is introduced. It is the affine parameter corresponding to the tangent vectors of the outgoing null
geodesics.  Near the horizon, we find that it behaves as
\begin{equation}
 \Lambda = \overline{\Lambda} + \halb \kappa n^2 + \mathcal{O}(n^3) \:,
\end{equation}
where $\overline{\Lambda}$ is value of $\Lambda$ on the horizon and it is constant with respect to $n$. If we take the $T\delta S$ part from \eqref{H_sur_gss_integral} we are able to write it as
\begin{equation}
\label{TdS_gss}
 T \delta S = \inv{16\pi} \int \dd^2 x_\parallel 2 \kappa \kzav{\sqrt{\overline{\sigma}} \delta F'_h + F'_h \delta \sqrt{\overline{\sigma}}} \:.
\end{equation}
It is clearly visible that our intention is to replace both variations in this formula with variation with respect to the affine parameter $\Lambda$. The relationships between variations follow as
\begin{equation}
 \delta\sqrt{\overline{\sigma}} = \inv{2\kappa} \sqrt{\overline{\sigma}} \sigma_2 \delta \Lambda \:,
 \qquad
 \delta F'_h = \frac{\partial F'}{\partial R} \frac{\partial R}{\partial n} \frac{\partial n}{\partial \Lambda} \delta \Lambda =
 \frac{F''_h}{\kappa} \kzav{\half{3} \sigma_3 + \widehat{R}_\parallel - \halb \sigma_2^2} \delta \Lambda \:.
\end{equation}
Using these formulae in \eqref{TdS_gss} yields
\begin{equation}
 T \delta S = \inv{16\pi} \int \dd^2 x_\parallel \hzav{
 2F''_h \sqrt{\overline{\sigma}} \kzav{\half{3} \sigma_3 + \widehat{R}_\parallel - \halb \sigma_2^2}
 + F'_h \sqrt{\overline{\sigma}} \sigma_2
 }\delta\Lambda\:,
\end{equation}
the second member can be substituted with $\sigma_2$ expressed from the left hand side of Einstein equation \eqref{Einstein_gss} results in an integral
\begin{align}
  T \delta S = \inv{16\pi} \int \dd^2 x_\parallel \sqrt{\overline{\sigma}} &\left\{
      F'_h R_\parallel
      + 16 \pi T^1_1 + \right. \nonumber \\      
      &\left.+ \hzav{F_h - F'_h \kzav{\overline{R}_\parallel - 2\sigma_2}}
      - 2 D^A \hzav{
          F''_h D_A \kzav{\overline{R}_\parallel - 2\sigma_2}
      } 
  \right\} \delta \Lambda \:. \nonumber \\
\end{align}
Stokes theorem \cite{Poisson:2009pwt} can be applied on the last term and since the horizon is a closed surface, this term will vanish. Analogically to the spherical symmetric case, the term originated from second derivative of $F'$ vanishes. The first term can be interpreted as variation of energy, analogically to the spherical case, it consists of the classical energy term multiplied by a factor of $F'$. The second member of the right hand side is a classical pressure term. The third term offers two interpretations as in the previous case. 
The first and straightforward interpretation is a generalized dark energy pressure
\begin{equation}
  T \delta S = \delta E + p \delta V + p_\lambda \delta V \:,
  \qquad
  p_\lambda = \frac{F_h - F'_h \kzav{\overline{R}_\parallel - 2\sigma_2}}{16\pi}\:,
\end{equation}
where the definition of the dark energy pressure is identical to the one in the spherical symmetric case.
With a little care we can take the second path and interpret the last term as an internal entropy term
\begin{equation}
  T \delta S = \delta E + p \delta V - T \delta S^i \:,
  \qquad
  \delta S^i = \frac{A}{4} \frac{F'_h \kzav{\overline{R}_\parallel - 2\sigma_2}  - F_h}{2\dot{N}} \delta \Lambda \:.
\end{equation}
Again, it is visible that the definition of the internal entropy reminds the one from the spherical case.

\end{document}